\documentclass[aps,twocolumn,superscriptaddress,floatfix,nofootinbib,showpacs,amsmath,amssymb,altaffilletter]{revtex4-1}
\pdfoutput=1
\usepackage{graphicx}
\usepackage{bm}
\usepackage{subfigure}
\usepackage{url}
\usepackage[hyperindex]{hyperref}
\usepackage{color}
\usepackage[ddmmyy,24hr]{datetime}
\usepackage{bigdelim}
\usepackage{booktabs}
\usepackage{dcolumn}
\usepackage{multirow}
\usepackage{subfigure}
\usepackage{cancel}
\usepackage{amsmath}
\usepackage{stackrel}
\usepackage{paralist}
\usepackage{xspace}
\usepackage[utf8]{inputenc}
\newcommand{\nua}[1]{\ensuremath{\rlap{\kern-2.5pt\ensuremath{\overset{\scriptscriptstyle(-)}{\phantom{\nu}}}}{\ensuremath{{\nu}_{#1}}}}}

\usepackage{xcolor}

\usepackage{soul}

\newcommand{\beq}{\begin{equation}}
\newcommand{\eeq}{\end{equation}}
\newcommand{\bea}{\begin{eqnarray}}
\newcommand{\eea}{\end{eqnarray}}

\usepackage{xcolor}

\usepackage{enumitem}

\makeatletter
\def\namedlabel#1#2{\begingroup
    #2%
    \def\@currentlabel{#2}%
    \phantomsection\label{#1}\endgroup
}

\makeatother
\begin{document}

\title{Incorporating the weak mixing angle dependence to reconcile the neutron skin measurement on $\mathbf{^{208}\mathrm{Pb}}$ by PREX-II}

\author{M.~Atzori Corona}
\affiliation{Dipartimento di Fisica, Universit\`{a} degli Studi di Cagliari,
Complesso Universitario di Monserrato - S.P. per Sestu Km 0.700,
09042 Monserrato (Cagliari), Italy}
\affiliation{Istituto Nazionale di Fisica Nucleare (INFN), Sezione di Cagliari,
Complesso Universitario di Monserrato - S.P. per Sestu Km 0.700,
09042 Monserrato (Cagliari), Italy}

\author{M.~Cadeddu}
\email{matteo.cadeddu@ca.infn.it}
\affiliation{Istituto Nazionale di Fisica Nucleare (INFN), Sezione di Cagliari,
Complesso Universitario di Monserrato - S.P. per Sestu Km 0.700,
09042 Monserrato (Cagliari), Italy}

\author{N.~Cargioli}
\email{nicola.cargioli@ca.infn.it}
\affiliation{Dipartimento di Fisica, Universit\`{a} degli Studi di Cagliari,
Complesso Universitario di Monserrato - S.P. per Sestu Km 0.700,
09042 Monserrato (Cagliari), Italy}
\affiliation{Istituto Nazionale di Fisica Nucleare (INFN), Sezione di Cagliari,
Complesso Universitario di Monserrato - S.P. per Sestu Km 0.700,
09042 Monserrato (Cagliari), Italy}

\author{P.~Finelli}
\affiliation{Dipartimento di Fisica e Astronomia, Università degli Studi di Bologna and INFN, Sezione di Bologna, Via Irnerio 46, I-40126 Bologna, Italy}

\author{M.~Vorabbi}
\affiliation{National Nuclear Data Center, Bldg. 817, Brookhaven National Laboratory, Upton, NY 11973-5000, USA}

\begin{abstract}\noindent
The only available electroweak measurement of the $^{208}\mathrm{Pb}$ neutron skin, $\Delta R_{\rm{np}}$, performed by the PREX-II Collaboration through polarized electron-lead scattering, shows a mild tension with respect to both the theoretical nuclear-model predictions and a host of measurements. However, the dependence on the weak mixing angle should be incorporated in the calculation, since its low-energy value is experimentally poorly known.
We first repeat the PREX-II analysis confirming their measurement by fixing the weak mixing angle to its standard model value. Then, we show the explicit dependence of the PREX-II measurement on the weak mixing angle, obtaining that it is fully degenerate with the neutron skin. To break this degeneracy, we exploit the weak mixing angle measurement from atomic parity violation on lead, obtaining a slightly thinner neutron skin but with about doubled uncertainties, possibly easing the PREX tension. Relying on the theoretical prediction, $\Delta R_{\rm{np}}^{\mathrm{th}}\approx 0.13$-$0.19\ \mathrm{fm}$, and using it as a prior in the fit, we find a weak mixing angle value about $1.2\sigma$ smaller than the standard model prediction.
Thus, we suggest a possible solution of the PREX-II tension by showing that, considering its underlying dependence on the weak mixing angle, the PREX-II neutron skin determination could be in agreement with the other available measurements and predictions if the weak mixing angle at the proper energy scale is smaller than the standard model prediction.
\end{abstract}
\maketitle

{\it Introduction}.---
The neutron skin, $\Delta R_{\rm{np}}\equiv R_n-R_p$, of a nucleus quantifies the difference between the neutron and the proton nuclear distribution radii, $R_n$ and $R_p$, respectively. Polarized electron-nucleus scattering happens through both the weak and the electromagnetic currents, therefore it provides an interesting way to assess the nuclear structure. Indeed, it is possible to isolate the weak interaction contribution, which strongly depends on the neutron nuclear density, from the electromagnetic one, which mainly probes the already highly-tested proton nuclear density.\\
Recently, the PREX-II Collaboration released their new measurement of the lead neutron skin through polarized electron-lead scattering, namely \mbox{$ \Delta R_{\rm{np}}^{\rm{PREX-II}}(^{208}\rm{Pb})=0.278\pm 0.078\ \mathrm{fm}$}~\cite{PhysRevLett.126.172502}, after a previous less precise measurement, $\Delta R_{\rm{np}}^{\rm{PREX-I}}(^{208}\mathrm{Pb})=0.30\pm 0.18\ \mathrm{fm}$~\cite{Abrahamyan_2012, Horowitz:2012tj}. The combination of these results leads to \mbox{$\Delta R_{\rm{np}}^{\rm{PREX,comb}}(^{208}\rm{Pb})=0.283\pm 0.071\ \mathrm{fm}$}~\cite{PhysRevLett.126.172502} indicating a preference for relatively large values of the skin. In particular, our analysis is focused on the PREX-II measurement, which is significantly more precise.\\ 
The PREX-II measurement 
is in tension with the other available determinations, such as those coming from electric-dipole polarizability~\cite{PhysRevC.85.041302,PhysRevC.92.064304,PhysRevC.104.024329}, antiprotonic atoms \cite{PhysRevLett.87.082501, Zhang:2021jwh,Klos:2007is}, proton-nucleus scattering \cite{PhysRevC.82.044611, PhysRevC.49.2118}, coherent pion photoproduction~\cite{PhysRevLett.112.242502} and the indirect measurements of neutron star observables~\cite{PhysRevLett.127.192701, PhysRevLett.120.172702, PhysRevLett.126.172503, PhysRevC.103.064323, Stone:2021uju, Essick:2021ezp, Riley:2021pdl,Miller:2021qha, Biswas:2021yge, LIGOScientific:2017vwq, LIGOScientific:2018cki,LIGOScientific:2020aai}. All these non-electroweak measurements are in fair agreement with each other, being also compatible with the predictions of different energy density functional (EDF) nuclear models~\cite{PhysRevC.85.041302,PhysRevC.92.064304,PhysRevC.104.024329}, $\Delta R_{\rm{np}}^{\rm{th}}(^{208}\rm{Pb})=0.13$-$0.19\ \mathrm{fm}$, and the first \textit{ab-initio} estimate of the lead neutron skin, $\Delta R_{\rm{np}}^{\textit{ab-initio}}(^{208}\rm{Pb})=0.14$-$0.20\ \mathrm{fm}$~\cite{hu2021ab}.
Although the hadronic probes have been shown to be affected by uncontrolled theoretical uncertainties~\cite{Thiel:2019tkm}, and the astrophysical constraints are rather indirect, the global picture that emerges indicates a preference for a thinner lead neutron skin with respect to PREX, which however is considered to be model independent being a non-hadronic probe.
So far, no solution to such a dilemma has been found, even if a reasonable compromise 
seems to be allowed for a specific set of quantified EDFs~\cite{Reinhard:2021utv}. 
Motivated by such a tension, we attempt to reconcile the PREX observation with all the other determinations by incorporating the so far neglected dependence on the weak mixing angle, $\sin^2\theta_W\equiv s^2_W$,  which is a key parameter of the electroweak theory~\cite{Zyla:2020zbs}.

{\it PREX-II analysis}.---
The PREX-II Collaboration measures the lead neutron skin exploiting 953 MeV polarized electrons scattering at forward angles on $^{208}\mathrm{Pb}$ nuclei~\cite{PhysRevLett.126.172502}. The polarization of the incoming electrons permits to isolate the weak-interaction contribution by building a parity violating asymmetry, $ \mathcal{A}_{\rm{pv}}$, as discussed in Ref.~\cite{PhysRevLett.126.172502}.
Adopting the plane-wave Born approximation, the asymmetry shows explicitly its dependence on the nuclear weak charge, $\mathcal{Q}_W$, and the weak, $F_W$, and charge, $F_{\rm{ch}}$, form factors
\begin{equation}
    \mathcal{A}_{\rm{pv}}\approx \frac{G_F Q^2}{4\pi \alpha \sqrt{2}}\frac{\mathcal{Q}_W F_W(Q^2)}{Z F_{\rm{ch}}(Q^2)},\label{APV_PREX-II}
\end{equation}
where  $G_F$ is the Fermi constant, $Q$ the momentum transfer, $\alpha$ the fine-structure constant and $Z$ the atomic number of the target nucleus. In particular, the weak and charge form factors represent the Fourier transforms of the corresponding weak and charge densities, $\rho_W$ and $\rho_{\rm{ch}}$.\\
The charge distribution of a lead nucleus has already been precisely tested through electromagnetic scattering processes, resulting in a charge distribution radius of $R_{\rm{ch}}(\mathrm{^{208}Pb})=5.503\pm 0.002 \ \mathrm{fm}$~\cite{DEVRIES1987495}. The charge distribution mainly depends on the proton nuclear density, so that $R_{\rm{ch}}$ can be translated into a determination of the proton root-mean-squared radius~\cite{PhysRevC.82.014320, Horowitz:2012tj}, leading to $R_{p}(\mathrm{^{208}Pb})=5.449\ \mathrm{fm}$~\cite{Horowitz:2012tj}.\\
However, the distorted-wave Born approximation (DWBA)~\cite{Uberall71,PhysRev.95.500,CIOFIDEGLIATTI1980163}, which incorporates the Coulomb distortion effects on the incoming electron approaching the target nucleus, must be adopted to properly describe  heavy nuclei, such as lead~\cite{Horowitz:1998vv}.\\
From Eq.~\ref{APV_PREX-II}, it is clear that PREX-II determines the weak form factor value at the experimental mean momentum transfer, $Q_\mathrm{PREX-II}\simeq 78\ \mathrm{MeV}$~\cite{PhysRevLett.126.172502}. After adopting a symmetrized two-parameter Fermi (2pF) function for the weak nuclear density~\cite{PhysRevC.94.034316},  it is possible to translate the weak form factor measurement into a determination of the weak nuclear radius, $R_W$.
In particular, the employed weak nuclear density reads~\cite{PhysRevC.94.034316}
\begin{equation}
    \rho_W (r,c,a)=\frac{3\ \mathcal{Q}_W}{4\pi c\ (c^2+\pi^2 a^2)}\dfrac{\sinh{(c/a)}}{\cosh{(c/a)}+\cosh{(r/a)}},\label{rhoWeak}
\end{equation}
where $r$ is the radius, $c$ represents the half density radius and $a$ is the so-called diffuseness, related to the thickness parameter $t$ through $t=4\,a\, \mathrm{ln}(3)$. We fixed the diffuseness parameter to the EDFs average value $a=0.605\pm0.025\ \mathrm{fm}$, as done by the PREX Collaboration in Ref.~\cite{PhysRevLett.126.172502}.
The corresponding weak charge radius $R_W$ can be obtained using~\cite{Horowitz:2012tj, PhysRevC.94.034316}
\begin{equation}
    R_W^2=\frac{1}{\mathcal{Q}_W}\int d^3r\, r^2 \rho_W(r,c,a)=\frac{3}{5}c^2 +\frac{7}{5}\pi^2 a^2,
\end{equation} which in turn can be
translated into the neutron distribution radius by the relation~\cite{Horowitz:2012tj}
\begin{equation}
    R_n^2=\frac{\mathcal{Q}_W}{\mathcal{Q}_W^n N}R_W^2-\frac{\mathcal{Q}_W^p Z}{\mathcal{Q}_W^n N}R_{\rm{ch}}^2-\langle r_p^2 \rangle-\frac{Z}{N}\langle r_n^2 \rangle + \frac{Z+N}{\mathcal{Q}_W^n N}\langle r_s^2\rangle,
\end{equation}
where $N$ is the neutron number, $\langle r_{p(n)}^2 \rangle$ is the charge radius of a single proton (neutron), namely, \mbox{$\langle r_{p}^2 \rangle=0.7071\ \mathrm{fm}^2$} and $\langle r^2_{n}\rangle=-0.1161 \ \mathrm{fm}^2$~\cite{Zyla:2020zbs}, $\langle r_s^2\rangle$ the nucleon strangeness radius ($\langle r_s^2\rangle \simeq -0.013\ \mathrm{fm}^2$~\cite{Horowitz:2018yxh}) and $\mathcal{Q}_W^{n, p}$ is the weak nucleon charges, defined as twice the opposite of the couplings of electrons to nucleons,
$g_{A V}^{e p}$ and $g_{A V}^{e n}$. Using the standard model (SM) prediction of the weak mixing angle at low energies, $s^{2\ \rm{SM}}_W=0.23857\pm0.00005$, and taking into account  radiative corrections~\cite{Zyla:2020zbs,Erler:2013xha,reverler}, one obtains $g_{A V,\text{SM}}^{e p}=- 0.0357$ and $g_{A V,\text{SM}}^{e n}=0.495$, where the coupling to protons is suppressed.
The weak charge of the nucleus, which enters directly both $\rho_W$ and $\mathcal{A}_{\rm{pv}}$, is the weak coupling to the nucleus, and it is defined as
\begin{align}
    \mathcal{Q}_W^{\mathrm{th}}(N, Z, s^2_W)
\equiv
- 2 [ Z g_{A V}^{e p}(s^2_W)
+ N g_{A V}^{e n} ] +\square^{\mathrm{Pb}}_{\gamma Z}
. \label{WeakChargeLead}
\end{align}
For $^{208}\mathrm{Pb}$, $Z$=82 and $N$=126. Thus, taking into account the radiative corrections and the total $\gamma$-$Z$ correction, $\square_{\gamma Z}$, for lead, the weak charge is predicted to be $\mathcal{Q}^{\mathrm{SM}}_W(^{208}\mathrm{Pb})=-117.9(3)$~\cite{PrivateCommunication} in the case of polarized electron scattering.\\ 
As shown in Eq.~\ref{APV_PREX-II}, PREX-II depends on the weak mixing angle at its experimental energy scale, where the latter is not well constrained. Indeed, despite the PREX-II measurement is said to be negligibly dependent on $s^2_W$~\cite{Donnelly:1989qs}, its dependence should be taken into account. \\
\begin{figure}[h]
    \centering
    \includegraphics[scale=0.55]{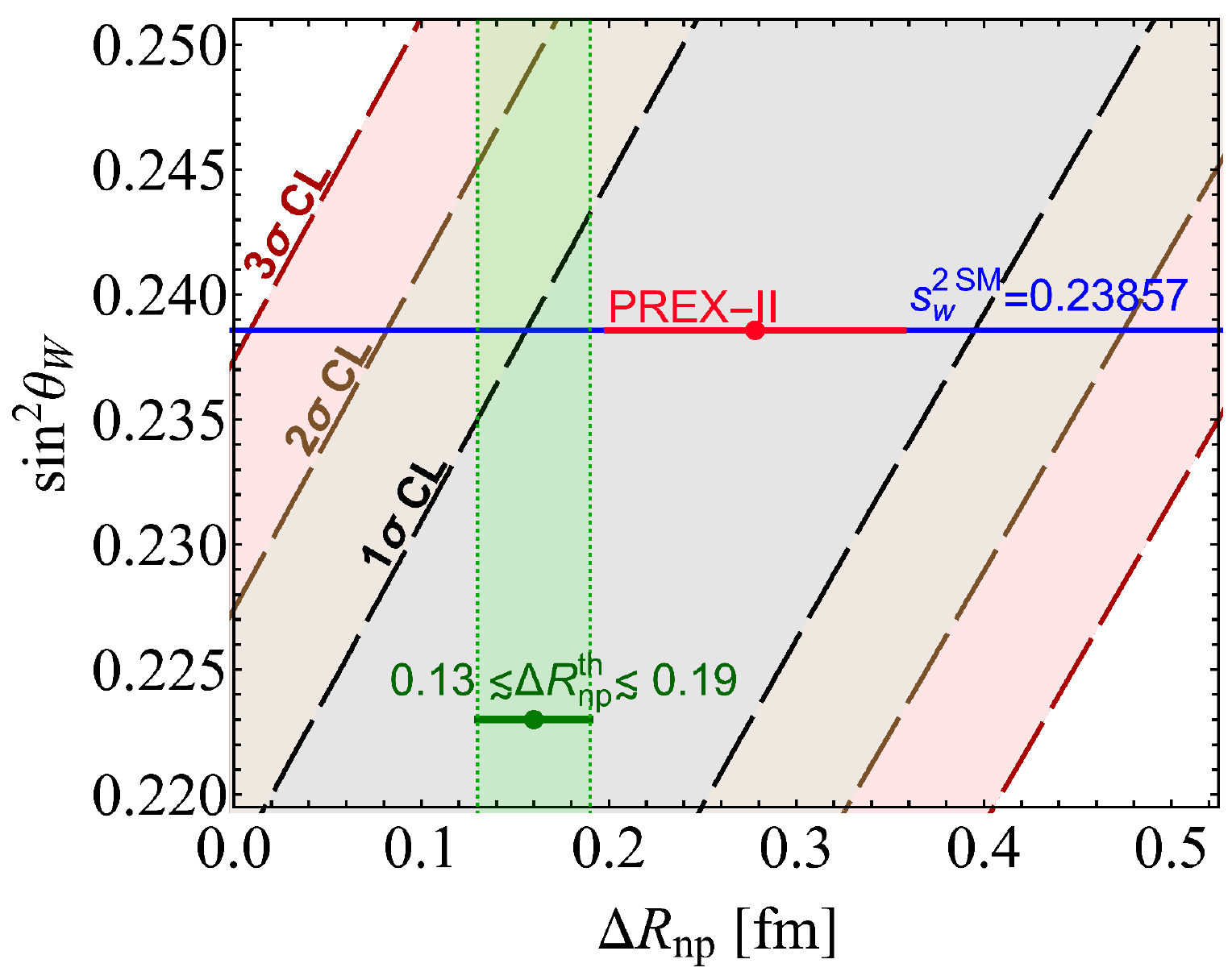}
    \caption{Favored regions in the $\sin^2\theta_W$ vs $\Delta R_{\rm{np}}(^{208}\mathrm{Pb})$ plane given by our reanalysis of the PREX-II data. The gray, brown and red shaded areas represent the 1, 2 and 3$\sigma$  confidence level contours. The red horizontal bar shows the PREX-II result~\cite{PhysRevLett.126.172502} for $s^{2\ \rm{SM}}_W$ (blue line). The green vertical band corresponds to the theoretical prediction, $\Delta R^{\rm{th}}_{\rm{np}}(^{208}\mathrm{Pb})$.}
    \label{fig:PREX-IIonly_2D}
\end{figure}

To do so, we modified the \textit{DREPHA} code~\cite{Drepha} to calculate the asymmetry and cross-section values under the DWBA, giving $\rho_W$ as an input evaluated for a variety of $R_W$ and $s^2_W$ values using the symmetrized 2pF parametrization~\cite{PhysRevC.94.034316}. Then, the asymmetries have been averaged considering the experimental angular acceptance $\epsilon(\theta)$, following the PREX-II procedure~\cite{PhysRevLett.126.172502}.
We fit the measured asymmetry $A_{\rm{pv}}^{\rm{meas}}=550\pm 16\ \mathrm{(stat.)}\pm 8\ \mathrm{(syst.)}\ \mathrm{ppb}$~\cite{PhysRevLett.126.172502}, computing the following $\chi^2$ function
\begin{equation}
    \chi^2_{\rm{PREX-II}}(s^2_W,R_n)=\Big(\frac{A_{\rm{pv}}(s^2_W,R_n)-A_{\rm{pv}}^{\rm{meas}}}{\sigma_{A_{\rm{pv}}^{\rm{meas}}}}\Big)^2 \,\label{chiSquare},
\end{equation}
where $\sigma_{A_{\rm{pv}}^{\rm{meas}}}$ is the total experimental uncertainty obtained summing in quadrature the statistical and systematical uncertainties. \\
We first perform a fit of the neutron skin fixing $s^2_W=s^{2\ \rm{SM}}_W$, leading to \mbox{$\Delta R_{\rm{np}}(s^{2\ \rm{SM}}_W)=0.276\pm0.078\ \mathrm{fm}$}, which confirms the PREX-II published result, $\Delta R_{\rm{np}}^{\rm{PREX-II}}(^{208}\mathrm{Pb})=0.278\pm 0.078\ \mathrm{fm}$~\cite{PhysRevLett.126.172502}, via an independent analysis.\\
Then, leaving $s^2_W$ free to vary, we obtain the result presented in Fig.~\ref{fig:PREX-IIonly_2D}.
The introduction of $s^2_W$ as a parameter in the fit produces a fully degenerate diagonal band. 
Therefore, smaller values of the neutron skin can be accessed for lower values of $s^2_W$, so that a neutron skin compatible with $\Delta R^{\rm{th}}_{\rm{np}}(^{208}\mathrm{Pb})$ can be obtained if $s^2_W\approx 0.225$.

{\it Atomic Parity Violation experiment on lead}.---
To break the degeneracy, we exploit the atomic parity violation (APV) experiment on $^{208}\mathrm{Pb}$, a weak-interaction process between the electrons and the nucleus, sensitive to the nuclear weak charge. Indeed, these probes represent the lowest energy determinations of $s^2_W$, being the momentum transfer of few MeVs~\cite{BOUCHIAT198373}, i.e., $Q_\mathrm{APV(\rm{Cs})}\sim 2.4\ \mathrm{MeV}$ for cesium atoms and $Q_\mathrm{APV(\rm{Pb})}\sim 8\ \mathrm{MeV}$ for lead atoms.
So far, the most precise APV measurement was performed on cesium and it is extracted from the ratio of the parity violating amplitude, $E_{\rm APV}$, to the Stark vector transition polarizability, $\beta$, and by calculating theoretically $E_{\rm APV}$~\cite{Zyla:2020zbs, Dzuba:2012kx, Bennett:1999pd}. 
The parity nonconserving nuclear-spin-independent part of the electron-nucleus interaction Hamiltonian~\cite{PhysRevA.93.012501} depends on  
$\rho_{W}(\mathbf{r})$. In the original calculation of $E_{\rm APV}$, a charge-density distribution has been used instead of $\rho_{W}(\mathbf{r})$ since the charge radius was better determined. To keep this into account, a so-called neutron-skin correction was applied to the final calculation using the proper density distribution~\cite{PhysRevC.46.2587, Pollock1999, James_1999, Horowitz2001,Viatkina}. 
For the amplitude calculation of APV(Pb), a uniformly charged ball density was assumed~\cite{PhysRevA.93.012501}, with $R_{\rm{ch}}(^{208}\mathrm{Pb})=5.5010\, \rm{fm}$~\cite{PhysRevA.93.012501}.\\
Once $E_{\rm APV}$ is obtained, the quantity 
\mbox{$R_{\rm {th}}= \left( {{\rm Im}\, E_{\rm APV}/ M_{1}} \right)_{\rm th.}$}
can be defined, where $M_{1}$ is the reduced electric-dipole transition of the magnetic-dipole operator for the $6 p^2\,^3P_{0} \rightarrow 6 p^2\,^3P_{1}$ transition relevant for lead. 
The theoretical calculated value is 
$R_{\rm {th}}=-10.6(4) \times 10^{-8}(-\mathcal{Q}^{\mathrm{APV(Pb)}}_{W}/N)$~\cite{PhysRevA.93.012501} while experimentally two measurements are available, $R_\mathrm{exp}=-9.86(12) \times 10^{-8}$~\cite{PhysRevLett.71.3442} and $R_\mathrm{exp}=-9.80(33) \times 10^{-8}$~\cite{Phipp_1996}, therefore we consider the experimental average value, $R_{\rm exp}^{\rm av.}$.
The lead weak charge is then obtained through the ratio between $R_\mathrm{exp}$ and $R_{\rm {th}}$, which gives 
$\mathcal{Q}^{\mathrm{APV(Pb)}}_{W}=-117(5)$~\cite{PhysRevA.93.012501}.\\ 
The neutron skin correction can be defined as
\begin{equation}
\delta E^\mathrm{n.s.}_\mathrm{APV}(R_n)= \left[ (N/\mathcal{Q}^{\mathrm{APV(Pb)}}_{W})\left(1-(q_n(R_n)/q_p)\right) E_\mathrm{APV} \right],
\end{equation}
where $q_{p(n)}$ are the integrals over the proton and neutron nuclear densities~\cite{PhysRevC.46.2587,Pollock1999,James_1999,Horowitz2001,Viatkina}, respectively, determined as discussed in detail in Refs.~\cite{Cadeddu:2018izq,PhysRevC.104.065502}. In the $q_{p}$ calculation, we employ the charge density used in the original determination~\cite{PhysRevA.93.012501}, while for $q_{n}$ calculation the symmetrized 2pF was adopted. \\
The weak charge measurement is then obtained through
\begin{equation}
   \mathcal{Q}_W^{\mathrm{APV(Pb)}}(R_{n})=-N R_{\rm exp}^{\rm av.}\Big(\dfrac{M_1}{\mathrm{Im}(E_{\rm{APV}}+\delta E_{\rm APV}^{\rm n.s.}(R_{n}))}\Big)_{\rm th.}.
\end{equation}
This procedure has already been exploited in the case of cesium~\cite{Cadeddu:2018izq,PhysRevC.104.065502}. 
\\
The APV measurement, contrary to PREX, is mainly sensitive to $s^2_W$ and only feebly on $\Delta R_{\rm np}$. We build the following $\chi^2$ function for APV(Pb)
\begin{equation}
    \chi^2_{\rm{APV}(\mathrm{Pb})}(s^2_W,R_n)=\Big(\dfrac{\mathcal{Q}^{\rm{th,APV}}_{W}(s^2_W)-\mathcal{Q}_W^{\rm{APV}(\mathrm{Pb})}(R_{n})}{\sigma^{\rm{APV}(\mathrm{Pb})}(s^2_W,R_n)}\Big)^2,\label{chiSquareAPVPb}
\end{equation}
where $\sigma^{\rm{APV}(\mathrm{Pb})}(s^2_W,R_n)$ is the total uncertainty.
The weak charge $\mathcal{Q}^{\rm{th},APV}_{W}$ for APV experiments is slightly different from the one given for polarized electron scattering in Eq.~\ref{WeakChargeLead}, due to different radiative corrections. In particular, adopting the description in Refs.~\cite{Zyla:2020zbs,PhysRevC.104.065502}, $\mathcal{Q}^{\rm{th},APV}_{W}=-118.79(5)$.
We perform a combined fit of the APV(Pb) and PREX-II measurements by summing the $\chi^2$ functions in Eqs. \ref{chiSquare} and \ref{chiSquareAPVPb} to fully exploit their correlations, requiring $s^2_W$ to be constant between the corresponding experimental momentum transfers, $8\lesssim Q\lesssim 78\ \mathrm{MeV}$.\\
\begin{figure}[h]
    \centering
    \includegraphics[scale=0.48]{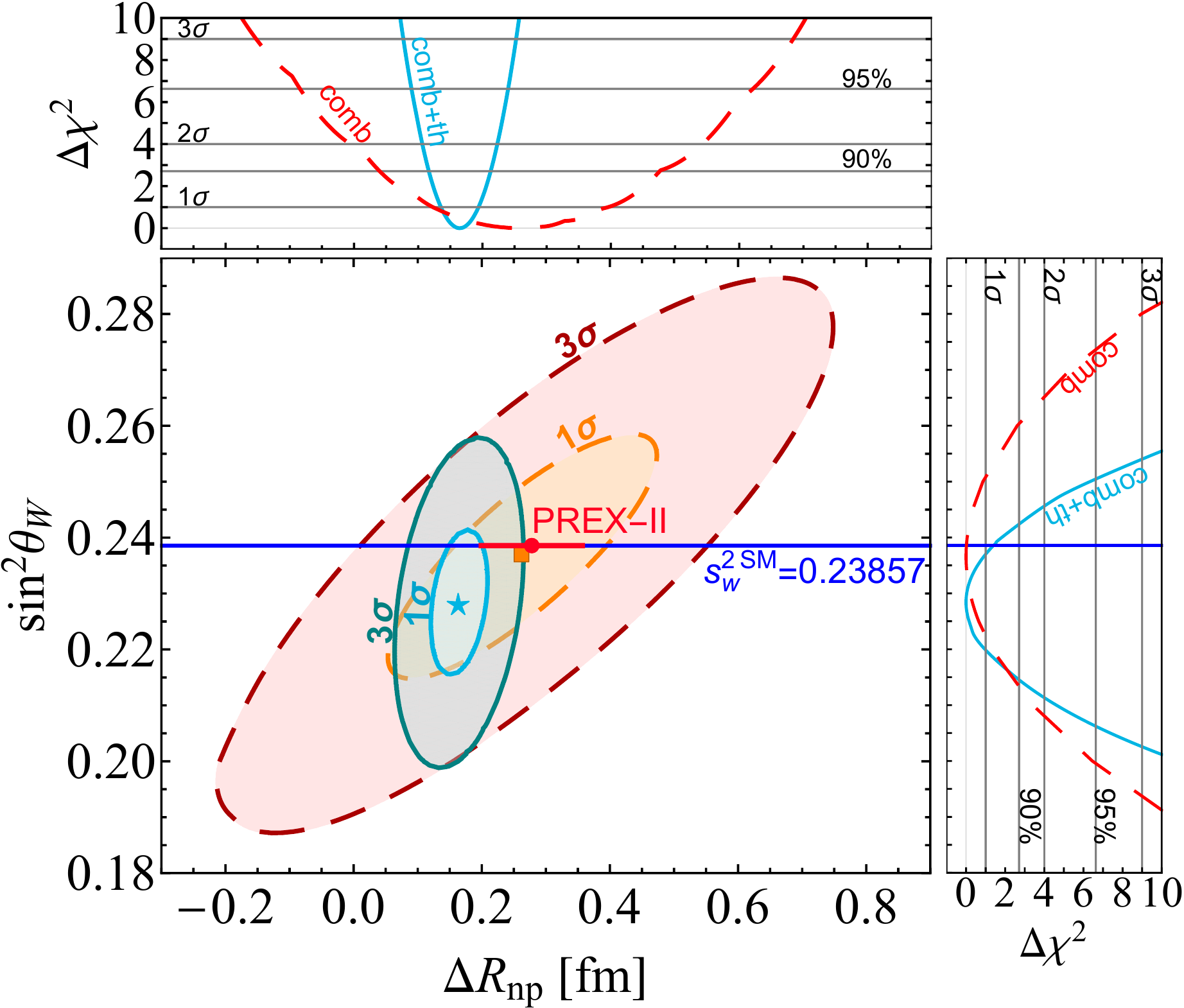}
    \caption{Combined fit results of APV(Pb)+PREX-II (dashed orange and dark red contours) and APV(Pb)+PREX-II+theory (solid cyan and blue contours, with their corresponding best fits (orange square and cyan star, respectively), shown in the $\sin^2\theta_W$ vs $\Delta R_{\rm{np}}(^{208}\mathrm{Pb})$ plane at 1$\sigma$ and 3$\sigma$ confidence levels. The side panels show the one-dimensional marginalizations (red line for APV(Pb)+PREX-II, cyan line for APV(Pb)+PREX-II+theory) for both the fits. The red horizontal bar shows the PREX-II  result~\cite{PhysRevLett.126.172502} for  $s^{2\ \rm{SM}}_W$(blue line).}
    \label{fig:PREX-IIAPV_Marg}
\end{figure}

In Fig.~\ref{fig:PREX-IIAPV_Marg}, we show the combined fit result through the dashed contours. In particular,  
the different sensitivities on $s^2_W$ and $\Delta R_{\rm np}$ of the two experiments allow us to find an elliptical closed region in the parameter space with best-fit values equal to
\begin{eqnarray}
\Delta R_{\rm{np}}^{\rm{Comb}}&=&0.262\pm 0.136\ \mathrm{fm},\\
\sin^2\theta_W(8\lesssim &Q&\lesssim 78\ \mathrm{MeV})=0.237\pm0.014.
\label{weakComb}
\end{eqnarray}
The neutron skin value obtained is slightly smaller than the PREX-II one~\cite{PhysRevLett.126.172502}, but with about double the uncertainty, so that now it is in agreement with $\Delta R_{\rm np}^{\rm th}$ as well as with other nonelectroweak measurements~\cite{PhysRevC.85.041302,PhysRevC.92.064304,PhysRevC.104.024329, PhysRevLett.87.082501, Zhang:2021jwh,Klos:2007is,PhysRevC.82.044611, PhysRevC.49.2118,PhysRevLett.112.242502,PhysRevLett.127.192701, PhysRevLett.120.172702, PhysRevLett.126.172503, PhysRevC.103.064323, Stone:2021uju, Essick:2021ezp, Riley:2021pdl,Miller:2021qha, Biswas:2021yge, LIGOScientific:2017vwq, LIGOScientific:2018cki,LIGOScientific:2020aai}. \\
From Fig.~\ref{fig:PREX-IIAPV_Marg}, it is clear that, to allow for thinner neutron skins, $s^2_W$ should be lower than its SM value. In Ref.~\cite{PhysRevC.104.065502}, this has already been observed for a combined fit of APV(Cs) and the coherent elastic neutrino-nucleus scattering data from the COHERENT experiment~\cite{Akimov:2021dab}. Surprisingly enough, looking at Fig. 3 of Ref.~\cite{PhysRevC.104.065502}, it can be retrieved that to obtain a cesium skin value around 0.13~fm, $s^2_W$ should be smaller than the SM prediction, in agreement with the results found in this work.\\
Finally, we combine the two experimental determinations with a Gaussian prior input on the neutron skin given by $\Delta R_{\rm{np}}^{\rm{th}}(^{208}\mathrm{Pb})=0.16\pm 0.03\ \mathrm{fm}$. 
The result of this combined fit is shown in Fig.~\ref{fig:PREX-IIAPV_Marg} through the solid contours.
Clearly, the prior forces the fit to favor smaller values of $s^2_W$. The best-fit values correspond to
\begin{eqnarray}
\Delta R_{\rm{np}}^{\rm{comb+\rm{th}}}&=&0.164\pm0.029\ \mathrm{fm},\\
\sin^2\theta_W(8\lesssim &Q&\lesssim 78\ \mathrm{MeV})=0.228\pm0.008.
\label{weakCombTh}
\end{eqnarray}
The $s^2_W$ best fit results in a value lower than the SM predicted one, and with smaller uncertainty with respect to the PREX-II+APV(Pb) combined fit result.\\

\begin{figure}[h]
    \centering
    \includegraphics[scale=0.55]{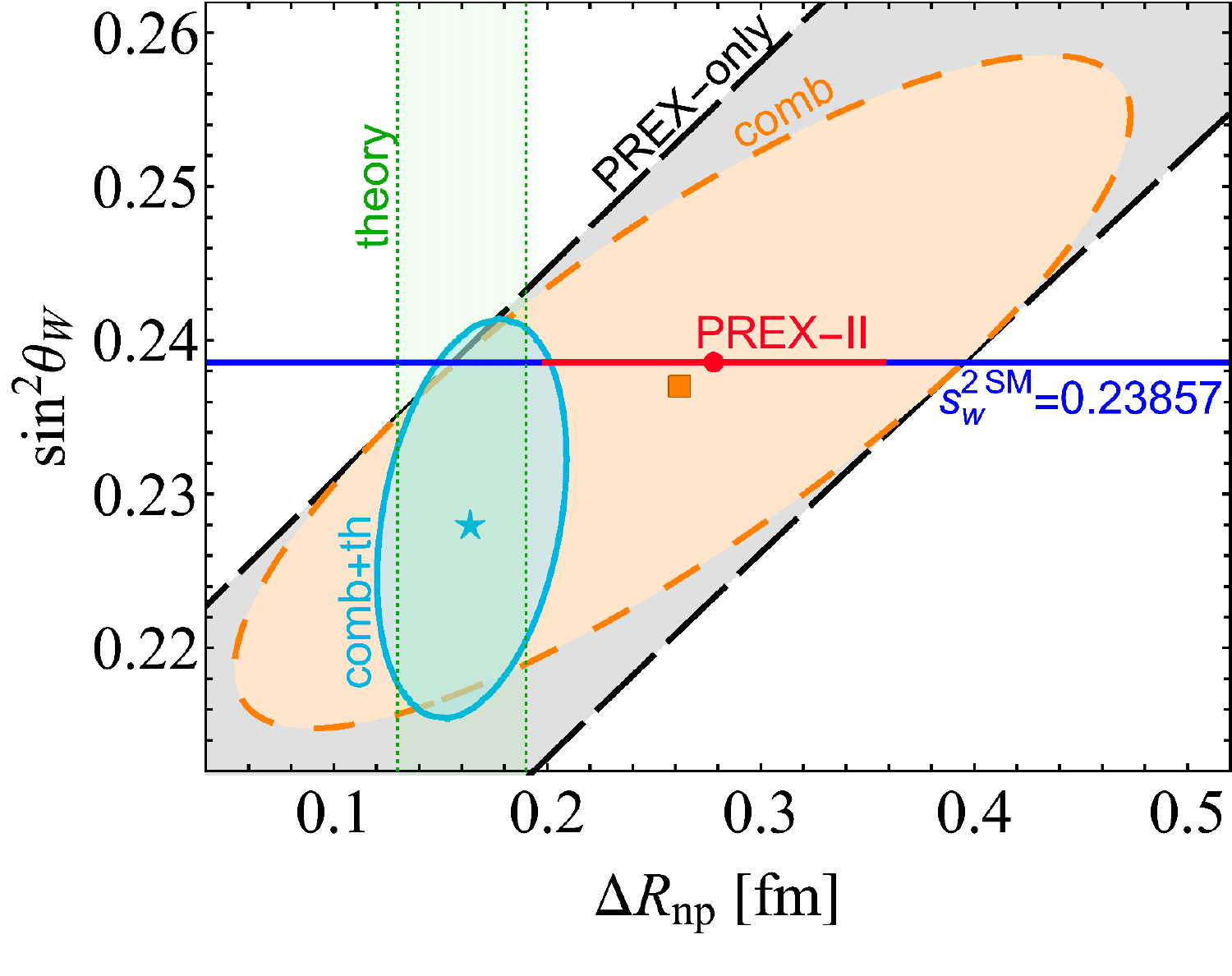}
    \caption{Summary of the PREX-only (grey long dashed), combined (orange dashed) and combined+theory (cyan solid) 1$\sigma$ confidence level contours in the $\sin^2\theta_W$ vs $\Delta R_{\rm{np}}(^{208}\mathrm{Pb})$ plane. The orange square and the cyan star points are the best fits of combined and combined+theory, respectively. The green vertical band shows $\Delta R_{\rm np}^{\rm th}$, while the red dot the PREX-II result~\cite{PhysRevLett.126.172502} for  $s^{2\ \rm{SM}}_W$(blue line).}
    \label{fig:SummurySkin}
\end{figure}

\begin{figure}[t]
    \centering
    \includegraphics[scale=0.57]{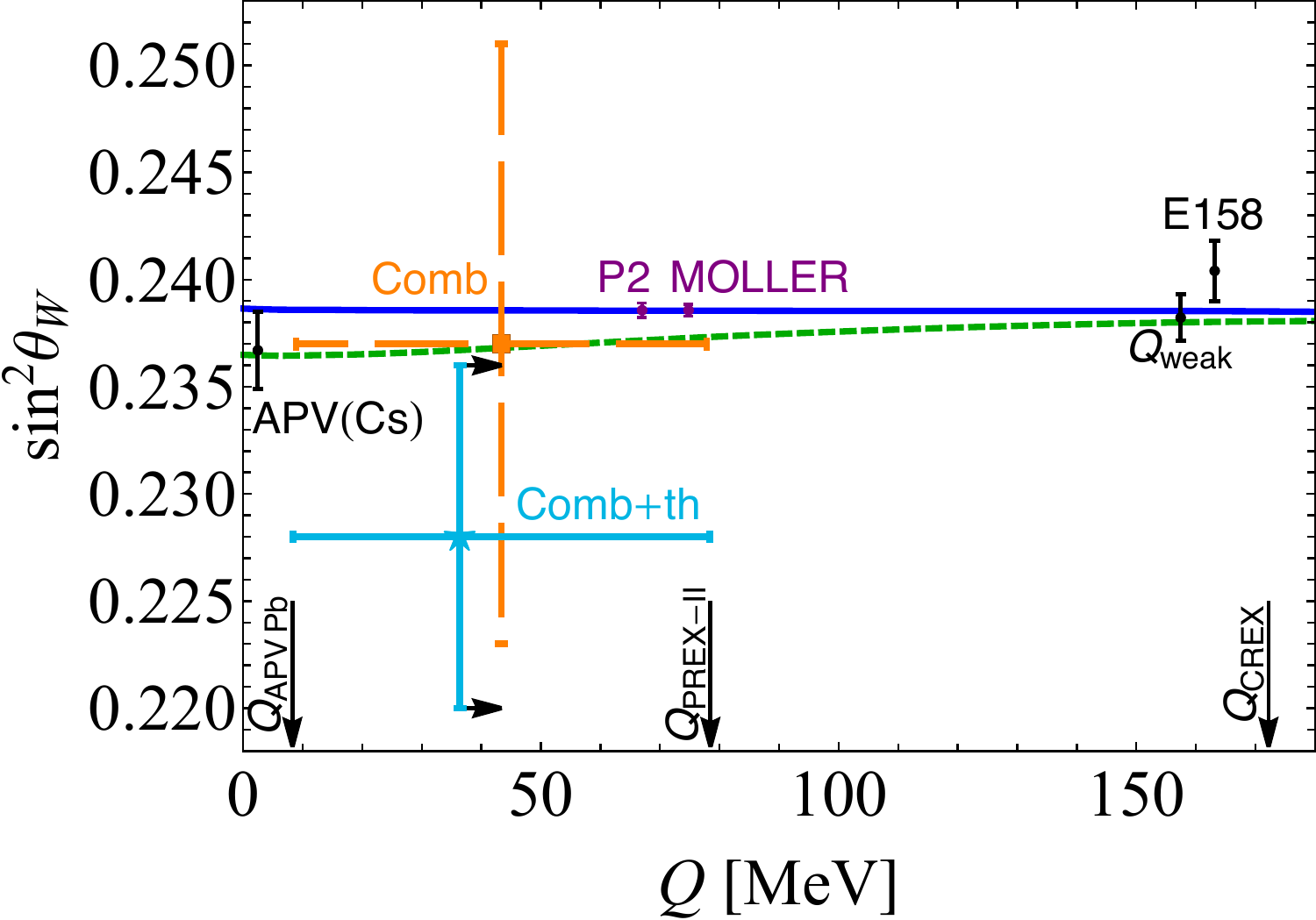}
    \caption{Weak mixing angle running with the energy scale $Q$. The SM prediction (solid blue curve) is compared with some experimental determinations (black dots)~\cite{Wood:1997zq,Dzuba:2012kx,Tanabashi:2018oca,Zyla:2020zbs,Androic:2018kni,Anthony:2005pm}, and future measurements (purple dots)~\cite{Becker:2018ggl,Dev:2021otb,Benesch:2014bas}.
    The orange dashed and the cyan solid points come from the combined and the combined+theory fits, respectively. The cyan result is shifted towards lower energies for illustrative purposes, as indicated by the arrows. The vertical arrows indicate the momentum transfer for APV(Pb), PREX-II and CREX, while the green dashed curve represents the modified running of $\sin^2\theta_W$ in a scenario involving a new mediator~\cite{Cadeddu:2021dqx}.}
    \label{fig:Running}
\end{figure}

{\it Main results and discussion}.---
The 1$\sigma$ confidence level contours obtained by the three presented analysis are summarized in Fig.~\ref{fig:SummurySkin} to underline how thinner skins are allowed for lower values of $s^2_W$.\\
Focusing on the implications for $s^2_W$, in Fig.~\ref{fig:Running} we summarize the state of the art of the weak mixing angle measurements in the low-energy regime ($Q\lesssim200\ \mathrm{MeV}$) through processes involving electrons.
The lowest energy determination belongs to APV(Cs), which is 1$\sigma$ lower than the SM value~\cite{Zyla:2020zbs}. The APV(Cs) value corresponds to a neutron skin correction determined from an extrapolation of neutron skin measurements from antiprotonic data~\cite{Dzuba:2012kx,PhysRevA.65.012106,PhysRevLett.85.1618,PhysRevLett.87.082501}, which is compatible with the EDF estimate on cesium. Let us note that the APV(Cs) result is currently debated in the community. The theoretical calculations have gone through many reevaluations, leading to different weak mixing angle determinations. For completeness, see the work presented in Ref.~\cite{Porsev:2009pr} and the recent calculation performed in Ref.~\cite{PhysRevD.103.L111303}.
At higher energies ($Q\approx 160\ \mathrm{MeV}$), the combined  $Q_{\rm{weak}}$~\cite{Androic:2018kni} and the E158~\cite{Anthony:2005pm} measurements precisely determine $s^2_W$ to be compatible with the SM prediction.
The orange square and the light blue star points are the results obtained in this work for the combined PREX-II+APV(Pb) fit, see Eq.~\ref{weakComb}, and the PREX-II+APV(Pb)+theory fit, see Eq.~\ref{weakCombTh}, respectively.
The horizontal error bars indicate that we assume $s^2_W$ to remain constant between the APV(Pb) and PREX-II experimental energy scales.

Since PREX-II and APV(Pb) are not so precise in measuring the weak mixing angle, future determinations at the same energy scale are awaited. In particular, the P2~\cite{Becker:2018ggl,Dev:2021otb} and the MOLLER~\cite{Benesch:2014bas} experiments are going to measure $s^2_W$ with high precision at energies slightly smaller than the PREX-II one.\\
The $s^2_W$ dependence on the energy scale is fundamental in our discussion. In fact, the presence of beyond the SM light particles could significantly modify the running of $s^2_W$ only at low-energies. The green dashed curve in Fig.~\ref{fig:Running} shows an example of a dark $Z$ boson~\cite{PhysRevLett.109.031802,PhysRevD.85.115019} of mass around $50\ \mathrm{MeV}$ as discussed in Ref.~\cite{Cadeddu:2021dqx}. It shows that $s^2_W$ can sensibly differ from the SM at low energies while remaining compatible for $Q\gtrsim150$ MeV, so that $Q_{\rm{weak}}$ and PREX-II could be measuring different $s^2_W$ values.\\
The PREX twin experiment on calcium, CREX~\cite{Mammei2013ProposalTJ,Kumar:2020ejz,Horowitz:2013wha}, is performed at $Q\approx170\ \mathrm{MeV}$, so near the $Q_{\rm{weak}}$ energy scale. In this regime, the value of $s^2_W$ is rather precisely measured to be close to $s^{2\ \rm{SM}}_W$, thus, if our interpretation is correct, CREX is expected to measure a thin calcium neutron skin, compatible with the prediction of coupled cluster calculations, $\Delta R_{\rm np}^{\rm cc}(^{48}\mathrm{Ca})=0.12$-$0.15\ \mathrm{fm}$~\cite{Hagen:2015yea}.\\

{\it Conclusions}.---
In summary, we confirm the PREX-II neutron skin measurement via an independent analysis at fixed weak mixing angle. Questioning the assumptions behind this measurement, we show the PREX-II neutron skin dependence on $s^2_W$, resulting in a fully degenerate band in the $s^2_W$ vs $\Delta R_{\rm{np}}(^{208}\mathrm{Pb})$ plane. Nevertheless, the fit suggests that a much smaller (or larger) lead neutron skin could be found if $s^2_W(Q_{\rm PREX-II})$ is small (or large) enough, so that neglecting such a dependence the experimental result might be misinterpreted. 
A combined fit with APV(Pb) allows us to to break this degeneracy and to obtain a closed favored region in the parameter space. The best fit is in great agreement with $s^{2\ \mathrm{SM}}_W$ and we find a slightly thinner neutron skin with respect to the PREX-II result, even if with double the uncertainty, showing that the PREX-II tension can be eased.
We find a smaller favored region in the parameter space by inserting $\Delta R_{\rm np}^{\rm th}$ as a prior, obtaining that a $s^2_W$ value 1.2$\sigma$ smaller than the SM one is needed to accommodate for smaller neutron skins.
Further $s^2_W$ measurements are needed, such as the upcoming P2 and MOLLER experiments~\cite{Becker:2018ggl,Dev:2021otb,Benesch:2014bas}, as well as new neutron skin determinations, such as the one by the MREX experiment~\cite{Becker:2018ggl}, to shed light on this tension.\\ 
In conclusion, we discuss the possibility that the upcoming CREX result could be only feebly affected by the $s^2_W$ dependence thanks to the higher experimental energy scale, where $s^2_W$ is experimentally constrained at the SM value, reconciling eventual conflicting results between the two twin experiments.

\begin{acknowledgements}
We would like to thank M. Gorchtein and  J. Erler for the fruitful discussions on the lead weak charge. Moreover, the useful discussions with F. Dordei, W. Bonivento and C. Giunti are kindly acknowledged.
The work at Brookhaven National Laboratory was sponsored by the Office of Nuclear Physics, Office of Science of the U.S. Department of Energy under Contract No. DE-AC02-98CH10886 with Brookhaven Science Associates, LLC.
\end{acknowledgements}

\bibliographystyle{apsrev4-1}
\bibliography{PREX}

\clearpage

\end{document}